\documentclass[twocolumn]{aastex631}

\begin{document}

\title{An Extragalactic Widefield Search for Technosignatures with the Murchison Widefield Array}

\correspondingauthor{Chenoa Tremblay}
\email{ctremblay@seti.org}

\author[0000-0002-4409-3515]{C.D. Tremblay}
\affiliation{SETI Institute, 339 Bernardo Ave, Suite 200, Mountain View, CA 94043, USA}
\affiliation{Berkeley SETI Research Center, University of California, Berkeley, CA 94720, USA}

\author[0000-0002-8195-7562]{S.J. Tingay}
\affiliation{International Centre for Radio Astronomy Research, Curtin University, Bentley, WA 6102, Australia}



\begin{abstract}
It is common for surveys that are designed to find artificial signals generated by distant civilizations to focus on galactic sources. Recently, researchers have started focusing on searching for all other sources within the field observed, including the vast population of background galaxies. Toward a population of galaxies in the background toward the Vela supernova remnant, we search for technosignatures, spectral and temporal features consistent with our understanding of technology. We set transmitter power limits for the detection of signals in a population of over 1,300 galaxies within a single field of view observed with the Murchison Widefield Array. 
\end{abstract}



\section{Introduction} \label{sec:intro}
When we consider the search for intelligent life beyond Earth, we often consider the age and advancement of technology that may produce a signal detectable by our telescopes. In popular culture, advanced civilizations are portrayed as having interstellar spacecraft and communications. Nikolai Kardashev \citep{Kardeshev_1964,1985IAUS..112..497K,1986pslu.book...25K}, and later modified by \cite{Gray_2020}, proposed the Kardashev scale to quantify the degree of technological advancement of intelligent life beyond Earth. The Kardashev scale has three levels: Type I civilizations are capable of accessing all the energy available on their planet (upward of 10$^{16}$\,W); Type II civilizations can directly consume a star's energy (upwards of 10$^{26}$\,W); and Type III civilizations can capture all the energy emitted by their galaxy (upward of 10$^{36}$\,W).

Civilizations on the upper end of the Kardashev scale could produce large quantities of electromagnetic radiation detectable at galactic distances. Some of the ideas explored in the past involve harnessing their galaxies' starlight \citep{Annis_99}, colonizing their solar system \citep{Gaviraghi_2016}, or using pulsars as communication networks \citep{CHENNAMANGALAM2015245,Haliki_2019}. The ability for radio waves to permeate space over long distances and through planetary atmospheres (e.g. \citealt{Tremblay_2022_SETI}) made them a practical method for searching for interstellar communication since \cite{Cocconi_1959} first proposed the concept. %No one knows at what frequency another civilization may broadcast to get attention. As Earth's earliest radio communication was at $\sim$144\,MHz frequency range \citep{Salisbury_1966}, Doppler motions caused by planets and the Galaxy, and the red-shifted signal over large distances provides motivation for searching low radio frequencies for extragalactic communication.}

\cite{Uno_2023} estimated the prevalence of extragalactic civilizations possessing a radio transmitter based on past $Breakthrough$ $Listen$\footnote{\url{https://breakthroughinitiatives.org}} observational data from \cite{Isaacson_2017} and \cite{Price_2020}. They determine the transmitter rate (TR) for four of the large surveys using the Parkes 64\,m (Murriyang) Telescope in New South Wales and the Robert C. Byrd Green Bank Telescope in West Virginia.  Based on an estimate of background galaxies in each field of view (FoV) searched they set upper limits of log(TR) ranging from 13.41 to 14.46, representing thousands of observational fields. Overall, they concluded that less than one in hundreds of trillions of extragalactic civilizations within 969\,Mpc possess a radio transmitter above 7.7~$\times$~10$^{26}$\,W of power, assuming one civilization per one-solar-mass stellar system.

\cite{Garrett_2023} and \cite{Choza_2024} represent other dedicated or opportunistic extragalactic searches with modern telescopes. Similar to \cite{Uno_2023}, \cite{Garrett_2023} searched for background galaxies within previously published fields from the Green Bank telescope. For this work, they reported a log(TR) of $\sim$10, with a sample of over 143,024 objects from 469 fields based on searches using NED and SIMBAD\footnote{\url{https://simbad.u-strasbg.fr/simbad/}}\citep{Wenger-SIMBAD}.  Their limits for the equivalent isotropic radiative power ranged from $\sim$10$^{23}$ to $\sim$10$^{26}$ depending on the object class.

With a large FOV, the Murchison Widefield Array (MWA; \citealt{Tingay_2013,Wayth_PhaseII}) in Western Australia provides a unique dataset to study galactic \citep{Tingay_2016,Tingay_2018,Tremblay_2020_SETI,Tremblay_2022_SETI}) and extragalactic communication. Currently, these frequencies (80--300\,MHz) are relatively unexplored for technosignatures, and at such a sensitivity that a Kardashev II or III civilization can be detected. As discussed by \cite{Garrett_2017}, aperture arrays like the MWA provide simultaneous coverage across up to 900 square degrees, containing thousands of galaxies and millions of stars. Using examples of low-frequency transmitters on Earth, we can consider the Air Force Space Surveillance System, known as `Space Fence,' which operated up until 2013. This system was a 1\,MW continuous wave (0.1\,Hz bandwidth) system operating at 216\,MHz. These factors add up to a powerful motivation for a study of the extragalactic population for technosignatures in this frequency range.

In this paper, we search for signals at a 10\,kHz spectral resolution originating from galaxies within a 400\,square\,degree FOV toward the Vela supernova remnant. This is the same field that was searched for signals from known stars and published by \cite{Tremblay_2020_SETI}. Here we explain the sample selection process and the results of this study, and discuss the results in the context of extragalactic communication and other searches of a similar nature.

\section{Observations}
The data for this analysis were obtained and processed as part of a molecular line survey as described by \cite{Tremblay_2020_NO}, but we provide some detail here. The data were collected between 2018 January 5 and 2018 January 23 using the MWA in a frequency range of 98--128\,MHz. For this dataset, 91 tiles (of 16 dipoles each) were online and baselines extended to 6\,km length. This resulted in a synthesized beam of 1\,arc\,minute full width at half maximum.

Each night of data was calibrated using observations of Hydra A and the calibration was further refined with phase-based self-calibration. Imaging of the data was done using {\sc WSClean} \citep{offringa-wsclean-2017} and flagged using {\sc aoflagger} \citep{Cotter_AOFlagger,OffriingaRFI}.  

After correction for ionospheric distortion and flux density scaling, each 5-minute observation was reference-frame corrected to the local-standard of rest. All the 5-minute observations that passed quality control tests were stacked together using inverse variance weighting for a total of 17 hours of integrated data time. The resultant data were a series of 24 three-dimensional cubes (each cube covering 1.28\,MHz of bandwidth) with 10\,kHz frequency resolution and a median root mean squared (RMS) noise of 0.035\,Jy\,beam$^{-1}$ for each 10 kHz channel, although the noise varied depending on the sky position and frequency. The largest impact of the frequency-dependent noise fluctuations was due to radio frequency interference (RFI) flagging in some observations compared to others. The sky-position dependence is a joint effect of the primary beam sensitivity pattern and positions of the sky where multiple processed fields had more data stacked\footnote{Even though the MWA can observe the entire sky, we only process a small segment of the field due to computational constraints. When imaging, some of the observations had slightly different phase centers, changing the sensitivity across the sky when all observations were stacked.} See Figure 3 in \cite{Tremblay_2020_NO} for an example of the sensitivity pattern on the sky.

\section{Source Selection}
\begin{figure*}
\includegraphics[width=0.48\textwidth]{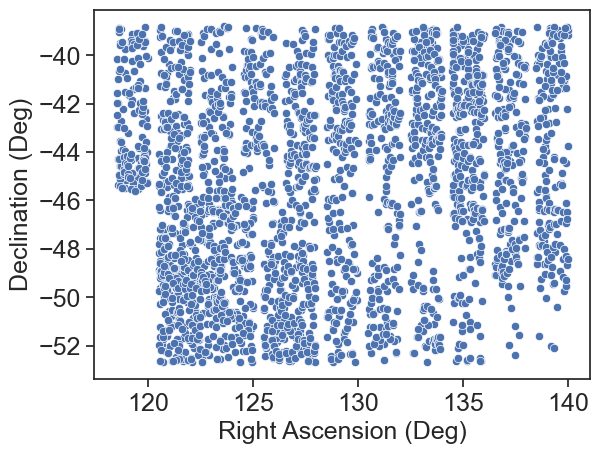}
\includegraphics[width=0.48\textwidth]{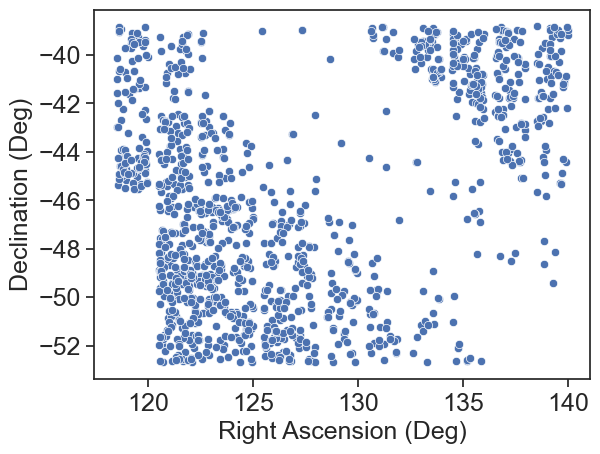}
\caption{Left: The distribution of the 2,880 galaxies found by searching through NED using the table access protocol (TAP) services available within TopCat \citep{Topcat}. Right: The distribution of 1,317 galaxies with known red shifts.}\label{fullset}
\end{figure*}

To complete a search for technosignatures from galaxies, we used the NASA/IPAC Extragalactic Database (NED) and searched for sources where the $pretype$ was listed as `G' for galaxy. Out of millions of objects cataloged in the 400 square degree field of view, there were a total of 2880 known galaxies from the full collection of surveys available in NED, including the Two Micron All Sky Survey (2MASS; \citealt{2MASS}), HI Parkes All Sky Survey (HIPASS; \citep{HIPPS}), \cite{Deep}, and \cite{ParkesHI} surveys (Figure 1). Of these galaxies, we focus on the 1317 sources that have known redshifts since these can be used to estimate distances, and hence transmitter power requirements. Fewer galaxies with known redshifts are found in regions where the Galactic plane extends through the field, as shown in the right-hand plot in Figure \ref{fullset}.

%and searched for where the $pretype$ was listed as 'G' for galaxy. Although millions of objects were identified in the field, in the 400 square degree FoV there were a total of 2,880 known galaxies from the full collection of surveys that NED has available. These catalogs included data from 2MASS \citep{2MASS}, HIPASS \citep{HIPPS}, \cite{Deep}, and \cite{ParkesHI} surveys. Of those, 1,317 have known red shifts and were the focus of our analysis, as approximate distances can be determined. As shown in Figure \ref{fullset}, there are gaps in the galaxy distribution where the Galactic Plane extends through the sky and few galaxies with known red shifts are in that region.

\section{Results}
\begin{figure*}
\centering
\includegraphics[width=0.48\textwidth]{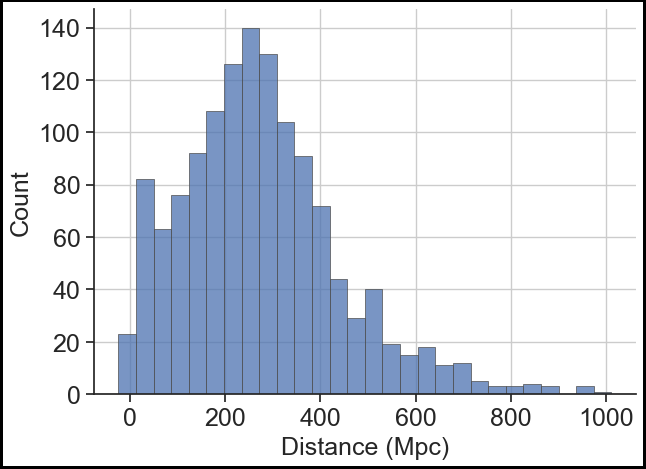}
\includegraphics[width=0.48\textwidth]{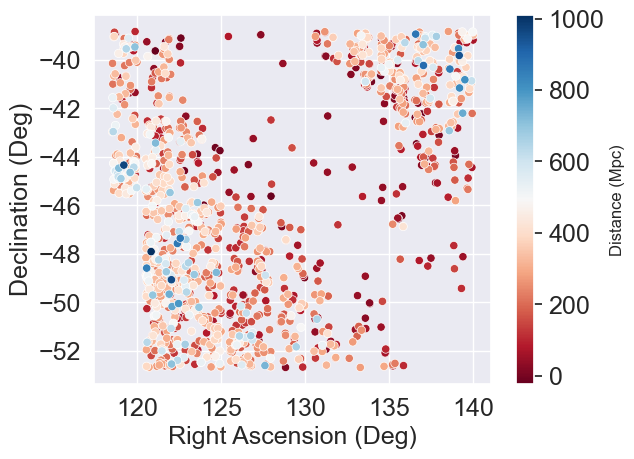}
\caption{Left:A histogram of the distances for the 1,317 galaxies with known red shifts in our sample. Right: A plot of the distribution of galaxies where the color represents the calculation of distance from Equation 1 converted to Mpc.}\label{distance}
\end{figure*}

Each of the 24 data cubes were searched blindly for any signal with a 6$\sigma$ value over local noise using {\sc Aegean} and with RMS maps created using {\sc BANE} (The Background and Noise Estimation tool) \citep{Hancock_Aegean}. No such signals were detected at this level or above. To determine the limits on transmitter power, we estimate the distance of the sources to the parsec scale from the red-shifts ($z$) reported for the galaxies in NED,

\begin{equation}
D=(c \times z)/H_o ~,
\end{equation}

where $D$ is the distance in parsecs, $c$ is the speed of light, and $H_{o}$ is the Hubble constant. However, we acknowledge that this might not be a definitive measure of distance as nearby galaxy red shifts will be dominated by random velocities rather than the Hubble flow (e.g. \citealt{Karachentsev_2009} determined a value of (78$\pm$2)\,kms$^{-1}$\,Mpc$^{-1}$ for galaxies less than 3\,Mpc from the Milky Way). We used the value of 67.4$\pm$0.5\,km\,s$^{-1}$\,Mpc$^{-1}$ \citep{Hubble_constant} for $H_{o}$ for the entire sample. The distances, shown in Figure \ref{distance}, range in values from 2.43$\times$10$^{1}$\,Mpc to 1.01$\times$10$^{3}$\,Mpc.

To determine the minimum detectable equivalent isotropic radiative power (EIRP$_{min}$) we use Equation 1 from \cite{Tremblay_2020_SETI},

\begin{equation}
\mathrm{EIRP}_{min} < 1.12 \times 10^{12}\, S_{rms}\, R^2 ~,
\end{equation}

where $S_{rms}$ is the local image noise in Jy\,beam$^{-1}$ and $R^2$ is the square of the distance in pc. For the nearest galaxies, we obtain an upper limit for EIRP$_{min}$ of 7.07$\times$10$^{22}$\,W with a maximum value of 4.35$\times$ 10$^{28}$\,W. The mean value for the survey is 4.40$\times$ 10$^{27}$\,W. We added the details regarding the closest 15 galaxies in Table \ref{tab1}.

\begin{figure}
\includegraphics[width=0.48\textwidth]{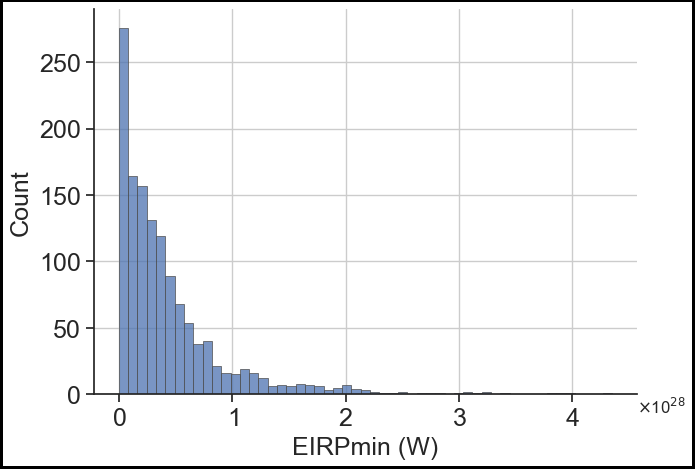}
\includegraphics[width=0.48\textwidth]{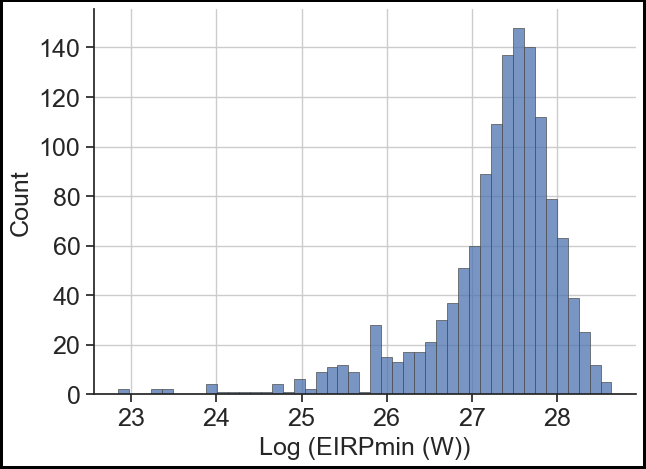}
\caption{A histogram plot of the EIRP$_{min}$ values calculated for this galaxy sample with the x-axis in Log (top) and Linear (bottom) space. }\label{EIRP}
\end{figure}

\begin{table*}
\caption{The nearest 15 galaxies and the isotropic limits.} 
\centering
\begin{tabular}{l c c c c c c c}
\hline \hline
Source Name	&	RA (deg)	&	Dec (deg)	&	Type	&	z	&	z reference	&	EIRPmin (W)	&	Distance (pc)\\
\hline
WISEA J091547.16-435623.7     	&	138.947	&	-43.940	&	G     	&	2.880E-04	&	\cite{Wen_EG}	&	7.068E+22	&	1.289E+06\\
WISEA J091153.11-430620.3     	&	137.971	&	-43.106	&	G     	&	3.110E-04	&	\cite{Wen_EG}	&	8.242E+22	&	1.392E+06\\
WISEA J090611.07-422218.5     	&	136.546	&	-42.372	&	G     	&	-4.640E-04	&	\cite{Wen_EG}	&	1.835E+23	&	-2.076E+06\\
WISEA J090331.66-413532.4     	&	135.882	&	-41.592	&	G     	&	-4.650E-04	&	\cite{Wen_EG}	&	1.842E+23	&	-2.081E+06\\
WISEA J085546.79-405336.0     	&	133.946	&	-40.893	&	G     	&	-5.260E-04	&	\cite{Wen_EG}	&	2.358E+23	&	-2.354E+06\\
WISEA J081419.97-435302.6     	&	123.583	&	-43.884	&	G     	&	-5.310E-04	&	\cite{Wen_EG}	&	2.403E+23	&	-2.376E+06\\
WISEA J090644.81-422939.0     	&	136.687	&	-42.494	&	G     	&	9.580E-04	&	\cite{Wen_EG}	&	7.820E+23	&	4.287E+06\\
WISEA J075810.20-411911.4     	&	119.543	&	-41.320	&	G     	&	9.630E-04	&	\cite{Wen_EG}	&	7.902E+23	&	4.309E+06\\
WISEA J081942.00-434437.3     	&	124.925	&	-43.744	&	G     	&	-1.002E-03	&	\cite{Wen_EG}	&	8.555E+23	&	-4.483E+06\\
WISEA J083949.90-485927.4     	&	129.958	&	-48.991	&	G     	&	1.025E-03	&	\cite{Wen_EG}	&	8.952E+23	&	4.586E+06\\
WISEA J081010.36-432230.8     	&	122.544	&	-43.375	&	G     	&	1.169E-03	&	\cite{Wen_EG}	&	1.164E+24	&	5.231E+06\\
WISEA J090629.45-432747.7     	&	136.623	&	-43.463	&	G     	&	-1.350E-03	&	\cite{Wen_EG}	&	1.553E+24	&	-6.041E+06\\
WISEA J090300.53-462604.6     	&	135.752	&	-46.435	&	G     	&	1.515E-03	&	\cite{Wen_EG}	&	1.956E+24	&	6.779E+06\\
WISEA J081114.35-443429.1     	&	122.810	&	-44.575	&	G     	&	-1.733E-03	&	\cite{Wen_EG}	&	2.559E+24	&	-7.754E+06\\
WISEA J081550.94-455942.4     	&	123.962	&	-45.995	&	G     	&	-2.211E-03	&	\cite{Wen_EG}	&	4.166E+24	&	-9.893E+06\\

\end{tabular}
\label{tab1}
\end{table*}

To compare these results to the surveys by \citep{Garrett_2023} and \citep{Uno_2023}, we use the transmitter rate calculation as per Figure 5 of \citep{Price_2020}:

\begin{equation}
\mathrm{TR} = (N_{\mathrm{star}}(\frac{\nu_{c}}{\nu_{tot}}))^{-1} ~,
\end{equation}

where $N_{\mathrm{star}}$ is the number of stars observed, $\nu_{\mathrm{c}}$ is the central frequency, and $\nu_{tot}$ is the total bandwidth. In this survey we cover a bandwidth of 30.72\,MHz, a central frequency of 113.28\,MHz, and 2,880 known galaxies. To determine the number of stars per galaxy we use the assumption made by \cite{Garrett_2023} that all galaxies contain the same number of stars at a value of $\sim$10$^{11}$. The log(TR) for our single FOV with the MWA is 15.06 if all galaxies are considered, or 14.69 if we determine the value for only those galaxies in which a distance was determined. 

\section{Discussion}
Although no signals were detected, we have placed stringent limits on the transmitter power over the 1~317 galaxies on the order of 10$^{22}$\,W. \cite{1985IAUS..112..497K} suggests that large-scale technology would have a lot of mass, a large energy potential, and a high information volume. With the sensitivity of modern radio telescopes, it may be possible to detect radiation coming from such a supercivilization even at galactic distances. 

\cite{Gaviraghi_2016} suggest methods in which a Kardashev Type II or Type III civilization may use significant power loads to colonize an entire solar system. They explore the ideas for strategies that range from noncrewed and crewed preparation phases or by sending mind-uploaded machines and the energy required for these ideas. Overall, they compute that the emitting power would be upwards of 10$^{24}$\,W, which is within the detection limits of this work.

It is generally agreed upon that it would be feasible for supercivilizations to harness the energy from natural objects and transmit signals over large distances \citep{Cirkovic_2015}. \cite{Haliki_2019} and \cite{CHENNAMANGALAM2015245} created a model where pulsars, rapidly spinning neutron stars, could be used as modulated signal beacons to contact other civilizations. Their recommendation was that Type III civilizations could not only use the beacons of light and change the modulation for communication but could potentially create an entire communication network. As modern radio telescopes can detect pulsars in galaxies external to our own, it offers another potential mechanism for generating emission, although broadband (frequency width over \,MHz) in nature; we could potentially detect the resultant energy from the technology.

With motivation of why these powerful signals may exist from supercivilizations, \cite{Choza_2024} presented results from a targeted search toward 97 nearby galaxies observed with the Green Bank Telescope. By comparing Figure \ref{EIRP} in this work with Figure 2 from \cite{Choza_2024}, we are up to 100 times less sensitive but cover a sample size that is more than 10 times larger, and still within the expected power limits of a Type II or Type III civilization. With the large FOV of the MWA, we also cover the entire galaxy with a single pointing and all 2,880 galaxies are observed in a single observation.

The limits on the EIRP$_{min}$ set for the MWA on this field are about 10 times less sensitive than \cite{Garrett_2023}, when similar distances are compared. Several close (red shifts $<$ 0.02) galaxies are included in their sample as well, where this sample extends to much greater distances.

Although both \cite{Uno_2023} and \cite{Garrett_2023} completed extragalactic searches in archive data, \cite{Garrett_2017} separated the limits based on galaxy classifications. This adds an additional level of analysis to technosignature searches when, as a community, we are trying to define where a signal may present itself. However, in all three studies (\citealt{Uno_2023}, \citealt{Garrett_2023}, and \citealt{Choza_2024}), they use single-dish telescopes with poor point source resolution. This MWA dataset has a resolution of 1\,arc\,min compared to the Green Bank telescope which has a 12\,arc\,min resolution at 1\,GHz. This means if a signal was detected it could be challenging to locate the exact source of emission.

\section{Conclusion}
This work represents the first low-frequency extragalactic technosignature search. Although the MWA is less sensitive when compared to the single-dish experiments, this work is important due to the unique frequency range coverage, large instantaneous FOV, and its point source resolution. Powerful emitters on Earth and some of our earliest radio transmissions are at frequencies less than 300\,MHz but only a few searches have been conducted.

We report EIRP$_{min}$ transmitter limits for 1,317 galaxies and searched toward over 2,880 galaxies for technosignatures. By setting a limit of $\sim$10$^{22}$\,W on a transmitter, we sit in a realm of possible detection based on theoretical models \citep{Gaviraghi_2016}. Although no artificial signals were detected, surveys of this kind are important for narrowing down the cosmic ``haystack" \citep{Wright}. %In the future, commensal surveys of this kind with the SKA will allow us to peer deeper into the cosmos with unprecedented sensitivity and increasing our chances of discovering something unknown.}

One of the main challenges in searching for signs of extraterrestrial intelligence is the speed at which we can observe the sky. Even with the large FOV offered by aperture arrays like the MWA, dedicated experiments done on shared-use instruments limit how much of the sky we cover and how often we cover the same sources. This landscape is significantly changing due to future and current commensal experiments on telescopes, such as the Karl G. Jansky Very Large Array (COSMIC; \citealt{Tremblay_COSMIC}) and the MeerKAT telescope (BLUSE; Czech et al., in preparation). Continuing to work together to cover the frequency space will be crucial in the future. %During its first year of operation, COSMIC observed over 800,000 sources, making it one of the largest SETI experiments in the world (Tremblay et al., in preparation). In terms of covering the search parameters for technosignatures, a commensal SETI experiment on SKA-Low and SKA-mid, with subarrays, wide bandwidth, and unprecedented sensitivity, would offer a significant leap forward.

\begin{acknowledgments}
C.D. Tremblay would like to thank Steve Croft and Carmen Choza for discussions and suggestions regarding this work. This research has made use of the NASA/IPAC Extragalactic Database (NED), which is operated by the Jet Propulsion Laboratory, California Institute of Technology, under contract with the National Aeronautics and Space Administration.

%\subsection {Facilities}
This scientific work makes use of the Murchison Radio-astronomy Observatory, operated by CSIRO. We acknowledge the Wajarri Yamatji people as the traditional owners of the Observatory site. Support for the operation of the MWA is provided by the Australian Government (NCRIS), under a contract to Curtin University administered by Astronomy Australia Limited.  Establishment of ASKAP, the Murchison Radio-astronomy Observatory, and the Pawsey Supercomputing Centre are initiatives of the Australian Government, with support from the Government of Western Australia and the Science and Industry Endowment Fund. 

\end{acknowledgments}


\vspace{5mm}
\facilities{MWA}


\software{The following software was used in the creation of the data cubes and data analysis:
{\sc aoflagger} and {\sc cotter} -- \cite{OffriingaRFI};
{\sc WSClean} -- \cite{offringa-wsclean-2014,offringa-wsclean-2017};
{\sc Aegean} -- \cite{Hancock_Aegean};
{\sc miriad} -- \cite{Miriad};
 {\sc TOPCAT} -- \cite{Topcat};
 {\sc Python PANDAS} -- \cite{reback2020pandas}
}



\bibliography{mwa-seti}{}
\bibliographystyle{aasjournal}



\end{document}